\documentclass{appolb}
\usepackage{graphicx}
\usepackage[utf8]{inputenc}

\usepackage[caption=false]{subfig}

\def\bbuildrel#1_#2^#3%
{\mathrel{\mathop{\kern 0pt#1}\limits_{#2}^{#3}}}
\newcommand{\g}{\gamma}
\newcommand{\bea}{\begin{eqnarray}}
\newcommand{\eea}{\end{eqnarray}}
\newcommand{\as}{a_s}
\newcommand{\G}{\Gamma}

\newcommand{\wtR}{\widetilde{R}}

\newcommand{\EQN}{\label}
\newcommand{\ovl}{\overline}
\newcommand{\unl}{\underline}
\newcommand{\ice}[1]{\relax}
\newcommand{\sbz}{  }
\newcommand{\nnb}{\nonumber}

\newcommand{\todoN}[1]{\relax}

\newcommand{\ed}{\end{document}}

\newcommand{\figB}{
{
\begin{figure}[b!]
\centering
\subfloat[]{
\hspace{.45cm}
\includegraphics[width=.85\linewidth,height=5.0cm]{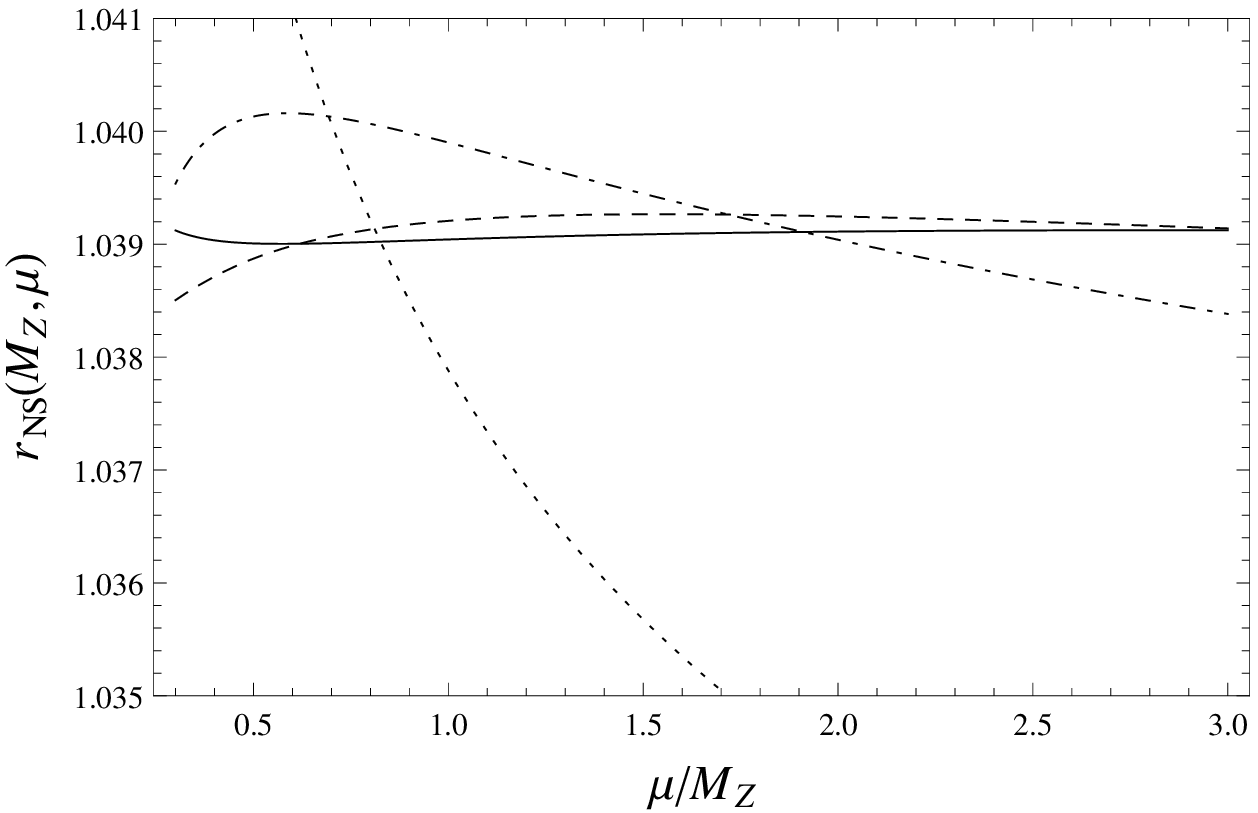}
\label{rns}
} \\
\subfloat[]{
\includegraphics[width=0.85\linewidth,height=5.9cm]{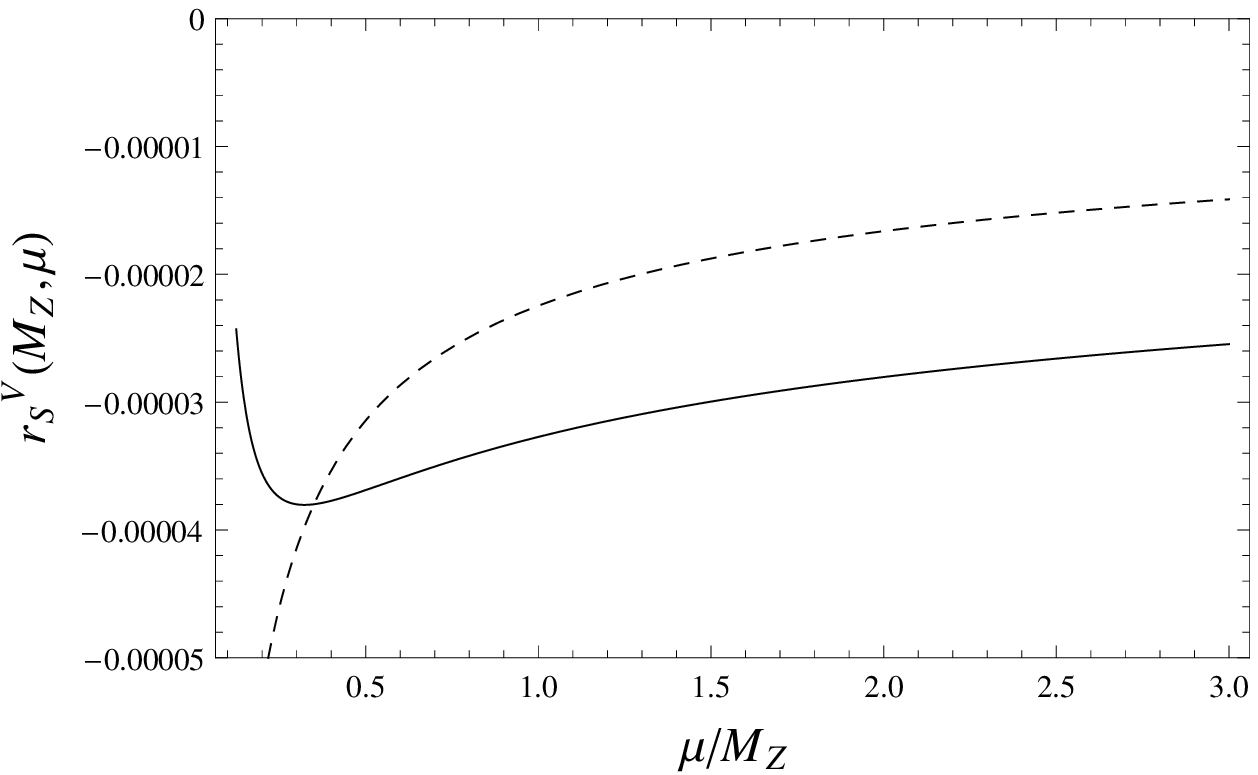}
\label{rsv}
} \\
\subfloat[]{
 \hspace{.2cm}
\includegraphics[width=.85\linewidth,height=5.0cm]{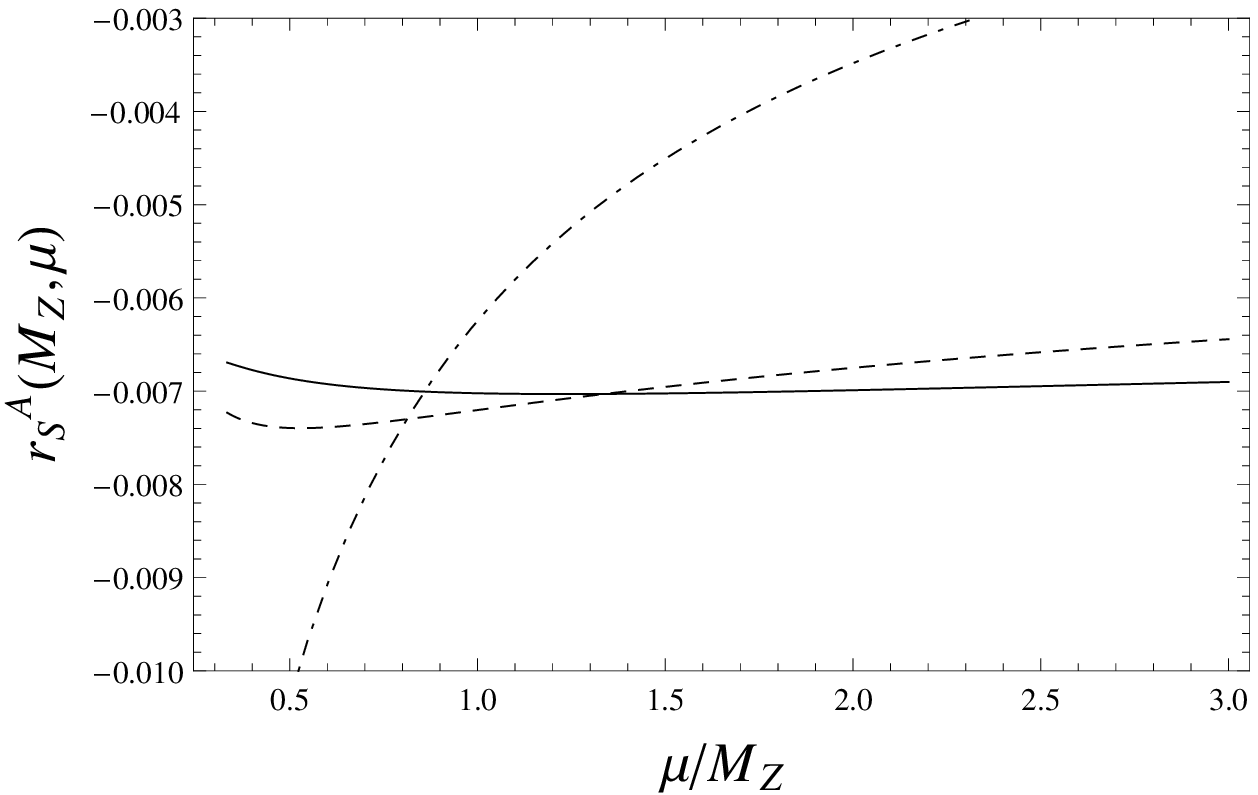}
\label{rsa}
}
\caption{
Scale dependence of (a) non-singlet $r_{NS}$ (b) vector singlet $r^V_S$ and (c) axial vector
singlet $r^A_{S;t,b}$. Dotted, dash-dotted, dashed and solid curves refer to ${\cal O}(\alpha_s)$
up to
${\cal O}(\alpha_s^4)$ predictions. $\alpha_s(M_Z)=0.1190$ and $n_l=5$ is adopted in all these
curves.
}
\label{Plots}
\end{figure}
}
}

\begin{document}
\title{Higgs Decay, Z Decay and the QCD Beta-Function\thanks{
Presented by J.H. K\"uhn at Matter To The Deepest,
Recent Developments In Physics Of Fundamental Interactions,
XLI International Conference of Theoretical Physics,
3-8 September 2017,
Podlesice, Poland}%
}
\author{P.A. Baikov
\address{Skobeltsyn Institute of Nuclear Physics, Moscow State University,
Moscow 119992, Russia}
\\ \vspace{5mm}
K.G. Chetyrkin
\\
\address{Institut f\"ur Theoretische Physik, Universit\"at Hamburg, 22761 Hamburg, Germany}
\\  \vspace{5mm}
J.H. K\"uhn
\address{Institute f\"ur Theoretische Teilchenphysik, Karlsruher Institut f\"ur Technologie (KIT), 76128 Karlsruhe, Germany}
}
\maketitle
\begin{abstract}
Recent developments in perturbative QCD, leading to the beta
  function in five-loop approximation are presented. In a first step the two
  most important decay modes of the Higgs boson are discussed: decays into a
  pair of gluons and, alternatively, decays into a bottom-antibottom quark
  pair. Subsequently the quark mass anomalous dimension is presented which is
  important for predicting the value of the bottom quark mass at high scales
  and, consequently, the Higgs boson decay rate into pair of massive quarks,
  in particular into $b\bar b$. In the next section the $\alpha_s^4$
  corrections to the vector-  and axial-vector correlator are discussed. These
  are the essential ingredients for the evaluation of the QCD corrections to
  the cross section for electron-positron annihilation into hadrons at low and
  at high energies,  to the hadronic decay rate of the
  $\tau$ lepton and for the $Z$-boson decay rate into hadrons.  Finally we
  present the prediction for the QCD beta-function in five-loop approximation,
  discuss the analytic structure of the result and compare with experiment at
  low and at high energies.
\end{abstract}

\section{Introduction}

During the past years significant progress has been made in the evaluation of
higher order QCD corrections to inclusive decay rates. Some of the basic tools
of these calculations have been formulated already long time ago (see, e.g.
early  reviews \cite{Harlander:1998dq,Steinhauser:2002rq}).

% \cite{Baikov:2015tea,Chetyrkin:2015mxa}).
 
\todoN{some of  Kostia's earlier papers on asymptotic freedom, e.g. arXiv:1701.08627 and
  others} 

However, steady progress has been achieved also more recently,
pushing e.g. the evaluation of QCD corrections to scalar- and vector-current
correlators to ${\cal O}(\alpha_s^4)$ and, correspondingly, the evaluation of
decay rates of scalar and vector particles to the same order 
\cite{Baikov:2005rw,Baikov:2008jh,Herzog:2017dtz}.
Also, along the same line the evaluation of the QCD beta function
has been pushed to fifth order \cite{Baikov:2016tgj}
and indeed also this result has been confirmed (and extended to  a generic  gauge  group)
by three new, independent
calculation \cite{Herzog:2017ohr,Luthe:2017ttg,Chetyrkin:2017bjc}.

\section{Dominant Higgs boson decay modes}

The two most important decay modes of the Higgs boson are the top quark mediated decay channel into 
two gluons and the decay into a bottom plus antibottom quark (for a recent review see \cite{Spira:2016ztx}). 
The higher order corrections to these 
modes have been evaluated up to ${\cal O}(\alpha_s^5)$ \cite{Baikov:2006ch,Davies:2017rle} for 
the two-gluon channel
(very recently even up to ${\cal O}(\alpha_s^6)$
\cite{Herzog:2017dtz}) and up to ${\cal O}(\alpha_s^4)$ for the $b\bar b$ channel \cite{Baikov:2005rw,Herzog:2017dtz}. 
Mixed terms related to  the $gg$ and the $b\bar b$ mode are treated in 
\cite{Davies:2017xsp}.
These two modes constitute the dominant Higgs decay channels with branching ratios around 15\% for 
the two-gluon and close to 60\% for the $b\bar b$ mode.

\subsection{Higgs decay into two gluons}

Let us start with the two-gluon channel. In the limit $m_t\to\infty$ the part of the effective 
Lagrangian which determines the coupling of the Higgs boson to gluons is given by
%Fig: top-triangle contracted to ggH coupling
\begin{equation}
%{\cal L}_{eff} = eqn 1 of arXiv:hep-ph/0604194
{\cal L}_{\rm eff}=-2^{1/4}G_F^{1/2}HC_1\left[O_1^\prime\right].
\end{equation}
Here $\left[O_1^\prime\right]$ is the renormalized counterpart of the bare operator 
\[O_1^\prime=G_{a\mu\nu}^{0\prime}G_a^{0\prime\mu\nu},\] 
with  $G_{a\mu\nu}$ standing  for the color 
field strength. The 
superscript $0$ denotes bare fields, and primed objects refer to the five-flavour effective theory. 
$C_1$ stands for the corresponding renormalized coefficient function, which carries all $M_t$ 
dependence. 
${\cal L}_{eff}$ thus effectively counts the number of heavy quark species, which in the Standard 
Model is restricted to the top quark. In Born approximation \cite{Inami:1982xt}
\begin{equation}
\Gamma_{\rm  Born}(H\to gg]=
\frac{G_FM_H^3}{36\pi\sqrt2}
\left(\frac{\alpha_s^{(n_l)}(M_H)}{\pi}\right)^2.
\end{equation}
The leading order result, being proportional to $\alpha_s^2$, exhibits a strong scale dependence 
which demonstrates the need for higher order corrections. Over the years subsequently higher orders 
have been calculated, from NLO 
through N${}^2$LO
\cite{Chetyrkin:1997iv} up to N${}^3$LO
\cite{Baikov:2006ch,Davies:2017rle}. Quite recently even the $N^4LO$ corrections became 
available \cite{Herzog:2017dtz}. (Power-suppressed corrections of order
$(m_H/M_t)^n$ with $n \le 5$  were calculated up to NNLO and can be found in
the literature \cite{Larin:1995sq,Schreck:2007um}.)

After a drastic increase of the cross section by about 60\% from the NLO corrections the N${}^2$LO 
terms lead to a further increase of the decay rate by about 20\%. This was the motivation for the 
evaluation of the N${}^3$LO terms. Using the optical theorem, the decay rate can be cast into the form
\begin{equation}
\Gamma(H\to gg)=
\frac{\sqrt{2}G_F}{M_H} C_1^2 {\rm Im} \Pi^{GG}(q^2=M_H^2) 
{},
\end{equation}
where
\begin{equation}
%\Pi^{GG}=
% from hep-hp/0604194
\Pi^{GG}(q^2) = \int \, e^{iqx} \langle 0|
T\left( 
\left[O_1^\prime\right](x)\left[O_1^\prime\right](0)
\right)
|0\rangle
{\rm\, dx}.
\end{equation}

The combination $\left[O_1^\prime\right]$ denotes the renormalized counterpart of 
the bare operator $O_1^\prime= G^{0'}_{a\mu\nu} G_a^{0'\mu\nu}$ and has been introduced above. The 
normalization $C_1$ is 
known to order $N^3O$ from massive tadpoles \cite{Chetyrkin:1997un}.

In total one finds
\begin{equation}
\Gamma(H\to gg) = \Gamma_{\rm Born}(H\to gg) \times K
\end{equation}
with
\begin{equation}
%K\to 1 + 17.9167 a_s' + complete
% from hep-hp/0604194
K = 1 + 17.9167 \, a'_s +   152.5 (a'_s)^2 +  381.5 (a'_s)^3.
% = 1 + 0.65038  + 0.20095   + 0.01825 {}.
\end{equation}
Here $a'_s=\alpha_s/\pi$. It is quite remarkable that the residual scale dependence is reduced quite 
drastically, from $\pm 24\%$ in LO to
$\pm 22\%$ in NLO down to $\pm 10\%$ in $N^2 LO$ and $\pm3\% $ in N${}^3$LO.

\subsection{Higgs decay into bottom quarks}

The second, and in fact dominant dominant decay mode of the Higgs boson is the $b\bar b$ channel. 
The decay rate into a quark-antiquark pair, generically denoted by $f\bar f$, is given by
\begin{equation}
%\Gamma (H\to f\bar f) = ...
% from hep-ph/0511063
\G(H \to \bar{f}f )
=\frac{G_F\,M_H}{4\sqrt{2}\pi}
m_f^2  \wtR (s = M_H^2)
\label{decay_rate_from_R}
{},
\end{equation}
where $\tilde R(s) = {\rm Im} \tilde \Pi(-s-i\epsilon)/(2\pi s)$ stands for
the absorptive part of the scalar two-point correlator
\begin{equation}
%\tilde \Pi(Q^2)=
%\tilde R (s)=
\EQN{Pi}
\widetilde{\Pi} (Q^2)
= (4\pi)^2 i\int dx e^{iqx}\langle 0|\;T[\;J^{\rm S}_f(x)
J^{\rm S}_{f}(0)\,]\;|0\rangle
{}.
\end{equation}
This five-loop result has been obtained in \cite{Baikov:2005rw} and recently
confirmed in \cite{Herzog:2017dtz,Luthe:2017ttg,Chetyrkin:2017bjc}.  Strong
cancellations are evident between ''kinematical terms'', originating from the
analytical transition from spacelike to timelike arguments, and ''dynamical
terms'', intrinsic for the calculation in the timelike region. In total one
finds
%\begin{equation}
%\tilde R= eq7 of PRL
% from hep-ph/0511063
\begin{eqnarray}
\widetilde{R} &=&  
{1}
{+} 
{5.667} \as
{+} a_s^2
%\left[
[
51.57 - \unl{15.63} 
- \, n_f (
 {1.907} 
- \unl{0.548} 
)
%zero == 0. + 0.*nf + 0.*x + 0.*nf*x
%\right]
]
\nonumber\\
&{+}& a_s^3
%\left[
[
{648.7} - \unl{484.6} 
- \, n_f ({63.74}  - \unl{37.97})
+ n_f^2 ({0.929}  \,
-\unl{0.67})
%zero == 0. + 0.*nf + 0.*nf^2 + 0.*x + 0.*nf*x + 0.*nf^2*x
%\right]
]
\nonumber\\
&{+}& \,a_s^4
%\left[
[
9470.8 - \unl{9431.4}
- n_f (1454.3   - \unl{1233.4})
\nonumber\\
&\phantom{+}&\phantom{+a_s^0}  \, +n_f^2 (54.78 - \unl{45.10})
- \, n_f^3 (0.454 -\unl{0.433})
%\right]
]
{},
\label{RSnumx}
\end{eqnarray}
%\end{equation}
where the underlined terms originate from the analytic continuation from the spacelike to the 
timelike region. Evidently the inclusion of the $\pi^2$ terms from higher orders alone does not 
improve the quality of the result. In total remarkable cancellations are observed between 
''kinematical'' and ''dynamical'' terms, leading to a nicely ''convergent'' answer. For $n_f=5$, 
the physically relevant result is given by 
\begin{eqnarray}
%\tilde R = formula
%         =numbers
%see eq5.4 of arXiv:1402.6611
%\hspace{-1cm}R^{S}
{\tilde R}(s = M_H^2,\mu=M_H) &=& 1 + 5.667\,  a_s+ 29.147 \, a_s^2  +
  41.758 \,a_s^3 \,  {- 825.7}\,a_s^4
\nonumber
\\
&=&
1 + 0.2041  + 0.0379  + 0.0020  {-0.00140}\, .
%\nonumber
\label{RS_as4_nl5}
\end{eqnarray}
In the last equation we have substituted $a_s(m_H)=\alpha_s/\pi =0.0360$, valid for a Higgs mass of 
125~GeV and $\alpha_s(M_Z)=0.118$. The nearly complete compensation between ${\cal O}(\alpha_s^3)$ 
and
${\cal O}(\alpha_s^4)$ term may be interpreted as a consequence of an accidentally small 
coefficient of the $\alpha_s^3$ term.

In total this leads to a dominant contribution of the $b\bar b$ mode with a branching ratio close 
to 60~\%. Note that an important ingredient in this context is the mass of the bottom quark at the 
scale of $m_H$, which has been taken as \cite{Herren:2017osy}
\todoN{chetykin+...?,\\ Herren:2017osy}
\begin{equation}
m_b(m_H)=2771 \pm 8|_{m_b} \pm 15|_{\alpha_s} {\rm MeV}\, .
\end{equation}
Let us mention in passing that, in order $\alpha_s^4$ there are also interference corrections 
resulting from mixed terms between $H\to gg$ and $H\to b\bar b$ which have been evaluated in 
\cite{Davies:2017xsp}.

\section{Quark mass anomalous dimension} 

It is well known that quark masses are conveniently defined to depend on a renormalization scale
\begin{equation}
%eq 11 of arXiv:1402.66
\mu^2\frac{d}{d\mu^2} {m}|{{}_{{g^0},
 m^0 }}
 = {m} \gamma_m(a_s) \equiv
-{m}\sum_{i\geq0}\gamma_{{i}}
\,
a_s^{i+1}
{},
\label{anom-mass-def}
\end{equation}
with $a_s=\alpha_s/\pi$ and the coefficients  $\gamma_i$  of the quark mass  anomalous dimension 
$\gamma_m$ are  known from $\gamma_0$ to 
$\gamma_4$ and thus in five-loop order \cite{Baikov:2014qja}. 
(At lower orders the  $\gamma_m$ was computed in 
\cite{Tarrach:1980up,Tarasov:1982gk,Larin:1993tp,Chetyrkin:1997dh,Vermaseren:1997fq}).
In numerical form and for $SU(3)$ 
it is given by
%\begin{equation}
%gamma_m for generic n as given in eq.4.1
%\end{equation}
\bea
\nonumber
&{}& \g_m = 
\nnb
\\
&-& a_s - a_s^2 \left(4.20833 - 0.138889 n_f\right)
\\ \nonumber
&-&
a_s^3  \left(19.5156 - 2.28412 n_f - 0.0270062 n_f^2 \right)
\\ \nonumber
&-&
a_s^4 \left(98.9434 - 19.1075 n_f + 0.276163 n_f^2  + 0.00579322 n_f^3 \right)
\\
&-&
a_s^5 \left(
559.7069 - 143.6864\, n_f + 7.4824\, n_f^2  + 0.1083\, n_f^3  - 0.000085359\, n_f^4
\right)
{}
\nonumber
%\label{N[gm5qcd]}
\eea
and, thus,
%\begin{equation}
%eq. 42 for specific n-values 3,4,5,6
%\end{equation}
\bea
\nonumber
\g_m \bbuildrel{=\!=\!=}_{n_f = 3}^{}
&-& \as - 3.79167 \,\as^2  - 12.4202 \,\as^3  - 44.2629 \,\as^4  - 198.907 \,\as^5
 ,\\ \nonumber
\gamma_m \bbuildrel{=\!=\!=}_{n_f = 4}^{}
&-& \as - 3.65278 \,\as^2  - 9.94704 \,\as^3  - 27.3029 \,\as^4  - 111.59 \,\as^5
,\\  \nonumber
\ \gamma_m \bbuildrel{=\!=\!=}_{n_f = 5}^{}
&-& \as - 3.51389 \,\as^2  - 7.41986 \,\as^3  - 11.0343 \,\as^4  - 41.8205 \,\as^5
 ,\\
\g_m \bbuildrel{=\!=\!=}_{n_f = 6}^{}
&-&  \as - 3.37500   \,\as^2  - 4.83867 \,\as^3  + 4.50817 \,\as^4  + 9.76016 \,\as^5
\nonumber
%\label{gm5:nf:3-6}
{}.
\eea

Note the significant cancellations between the contributions for $n_f^0$ and $n_f^1$ for values of 
$n_f$ around 4 and 5 which are clearly visible for the four-loop result and persist in five-loop 
order. This leads to a moderate growth of the series, even for scales as small as 2~GeV, where 
$a_s\equiv \alpha_s/\pi\approx 0.1$. The strong cancellations between different powers of $n_f$ 
have been anticipated by predictions based on ''Asymptotic Pad\'e Approximants'' 
\cite{Ellis:1997sb,Elias:1998bi,Kataev:2008ym}, the numerical value of the result, however, differs 
significantly (see table 1).

%Table 1 of arXiv:1402.6611

\begin{table}
\begin{center}
  \begin{tabular}{| c | c | c|c|c| }
    \hline
    $n_f$               &  3   & 4  &  5  &     6   \\ \hline
    $(\g_m)_4^{\rm exact}$       &  198.899   &  111.579  &  41.807  &     -9.777 \\ [0.51mm]    
\hline
    \rule{0mm}{5.5mm}
    $(\g_m)_4^{\rm APAP}$  \cite{Ellis:1997sb}         &  162.0   & 67.1  &  -13.7  &   -80.0 \\[1mm] 
    $(\g_m)_4^{\rm APAP}$  \cite{Elias:1998bi}         &  163.0   & 75.2  &  12.6  &   12.2 \\[1mm]  
    $(\g_m)_4^{\rm APAP}$  \cite{Kataev:2008ym}         &  164.0   & 71.6 &  -4.8  &   -64.6 \\[1mm] 
    \hline
  \end{tabular}
\end{center}
\caption{The exact results for $(\g_m)_4$ together with the predictions made with the help of
the original APAP method and   its two somewhat  modified versions.}
\end{table}

Let us note in passing that quite recently the result for a general
gauge group has been obtained
\cite{Luthe:2016xec,Baikov:2017ujl}.

\section{$Z$ decay in ${\cal O}(\alpha_s^4)$}

In view of asymptotic freedom perturbative QCD can be applied at vastly different energy scales, 
despite the dramatic variation of the strong coupling between the mass of the $\tau$ lepton and, 
for example, the mass of the $Z$ boson. Starting, for example, at the scale of the $\tau$-lepton 
with
\begin{equation}
\alpha_s(m_{\tau})= 0.332 \pm 0.005|_{exp} \pm 0.015|_{th}
\end{equation}
four loop running and matching at the flavour thresholds leads to the reduction of the strong 
coupling at the scale of the $Z$ boson mass
\cite{Baikov:2008jh}
\begin{equation}
\alpha_s(M_Z)= 0.1202 \pm 0.006|_{exp} \pm 0.0018|_{th} \pm 0.0003|_{evol} \end{equation}
by a factor three and a reduction of the uncertainties by nearly a factor ten. In this case the 
evolution error receives contributions from uncertainties in the charm- and bottom-quark mass, the 
variation of the matching scale and the four-loop truncation of the renormalization group equation. 
The final result is in remarkable agreement with the direct determination of $\alpha_s$ from $Z$ 
decays which leads to
$\alpha_s= 0.1190 \pm 0.0026|_{exp}$ and a small theory error. Note that the dominant term in the 
QCD corrections for $Z$ decays is identical to the correction term for $\tau$ decays.
However, starting from ${\cal O}(\alpha_s^2$), one receives additional,
new terms in the $Z$ boson case. These arise from so called singlet contributions which in turn 
are different for the vector and the axial vector part.

%plot FIG 1 of 1201.5804

\begin{figure}[htb]
%\begin{figure}[b!]
%\begin{figure}[H]
%\centering
\subfloat[]{
\includegraphics[width=.25\linewidth]{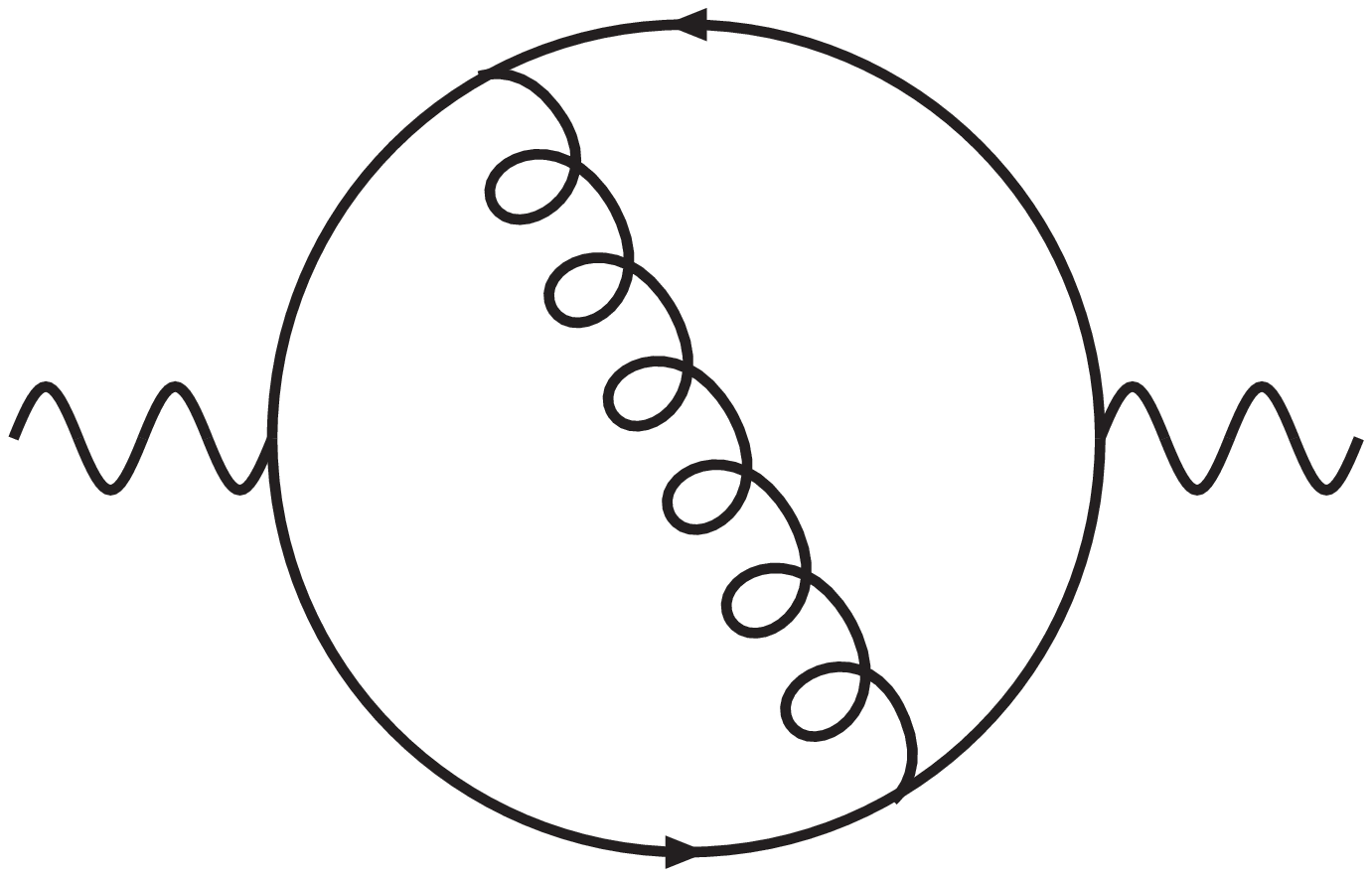}
\label{PropNS}
}
\subfloat[]{
 \includegraphics[width=.3\linewidth]{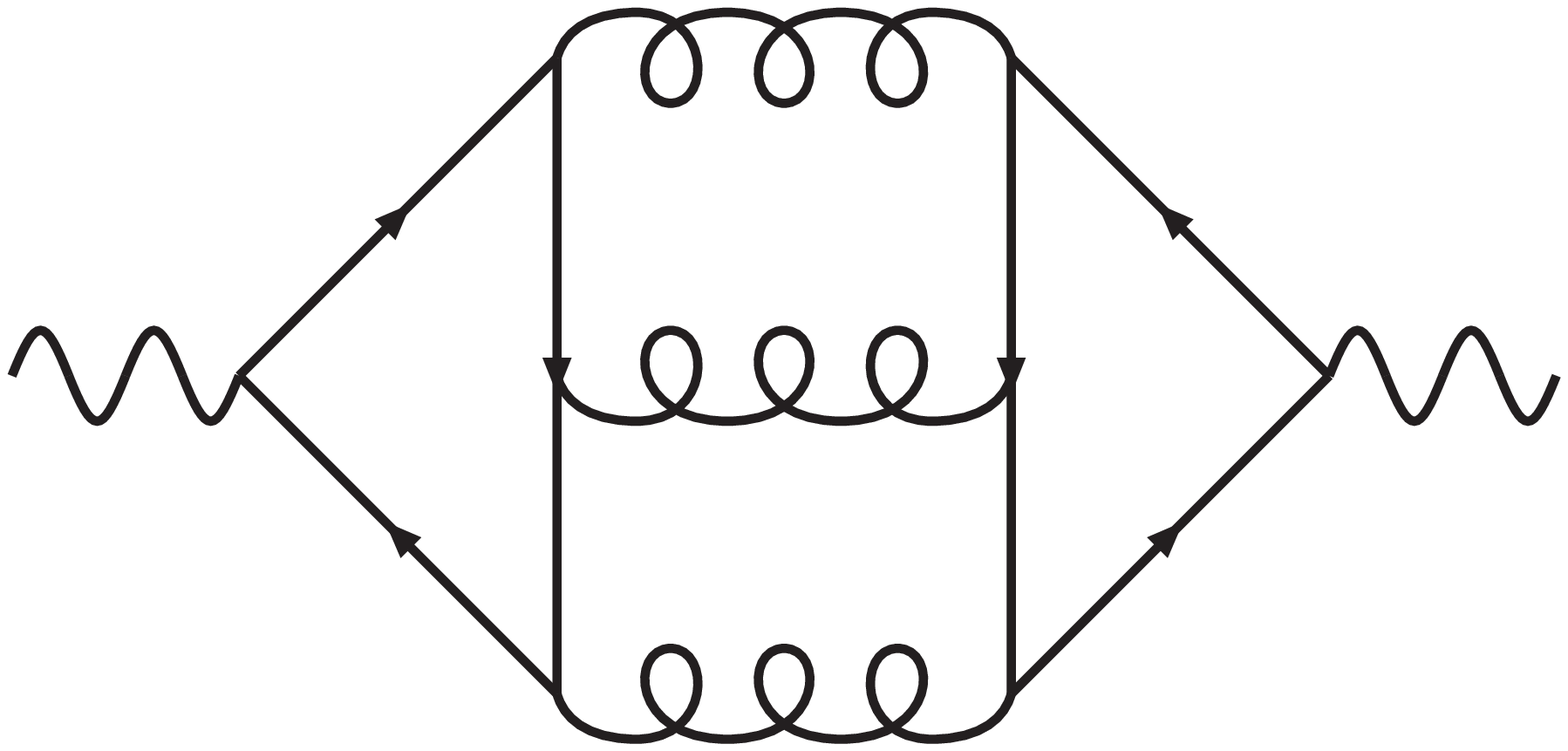}
\label{PropSV}
}
\subfloat[]{
 \includegraphics[width=.3\linewidth]{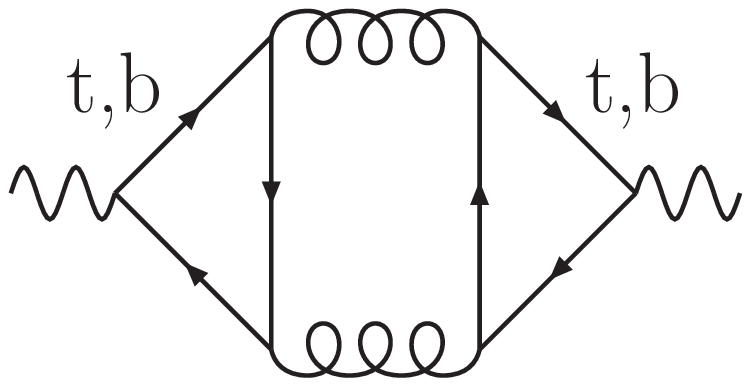}
\label{PropSA}
}
\caption{Different contributions to $r$-ratios: (a) non-singlet, (b) vector singlet and (c) axial 
vector singlet.}
\end{figure}

\figB
\clearpage

In total one finds for the QCD corrected decay rate of the $Z$ boson
(neglecting for the moment mass suppressed terms of ${\rm O}(m_b^2/M_Z^2)$
and electroweak corrections)
\begin{equation}
% eq2 of 1201.5804 R^{nc}= ...
R^{\rm  nc}=3\Big[\sum_f\!v_f^2 r_{\rm NS}^V + \big(\sum_f\!v_f\big)^2 r_{\rm S}^V
+\sum_f\!a_f^2 r_{\rm NS}^A + r_{\rm S;t,b}^A\Big]~.
\end{equation}

The relative importance of the different terms is best seen from the results of the various 
$r$-ratios introduced above. In numerical form  \cite{Baikov:2012er}
%\begin{equation}
\bea
% eq.3 of 1201.5804
%r_{NS}=
%r^V_s=
%r_S^A=
%r^A_{s,tb}
%\begin{align}
 r_{\rm NS}&=&1 + \as + 1.4092\,\as^2 - 12.7671\,\as^3 - 79.9806\,\as^4~, \nnb \\
 r_{\rm S}^V&=&-0.4132\,\as^3 - 4.9841\,\as^4~, \nnb \\
 r_{\rm S:t,b}^A &=&  (-3.0833+l_t)\,\as^2  + (-15.9877+3.7222\,l_t+1.9167\,l_t^2)\,\as^3 \nnb \\
 \phantom{r_{\rm S:t,b}^A} &\phantom{=}& +( 49.0309-17.6637\,l_t+14.6597\,l_t^2 + 3.6736\,l_t^3)\,\as^4  ~,
%\end{align}
%\end{equation}
\eea
%
%{Fig.2 of 1201.5804}
%\newpage
\ice{
\begin{figure}[b!]
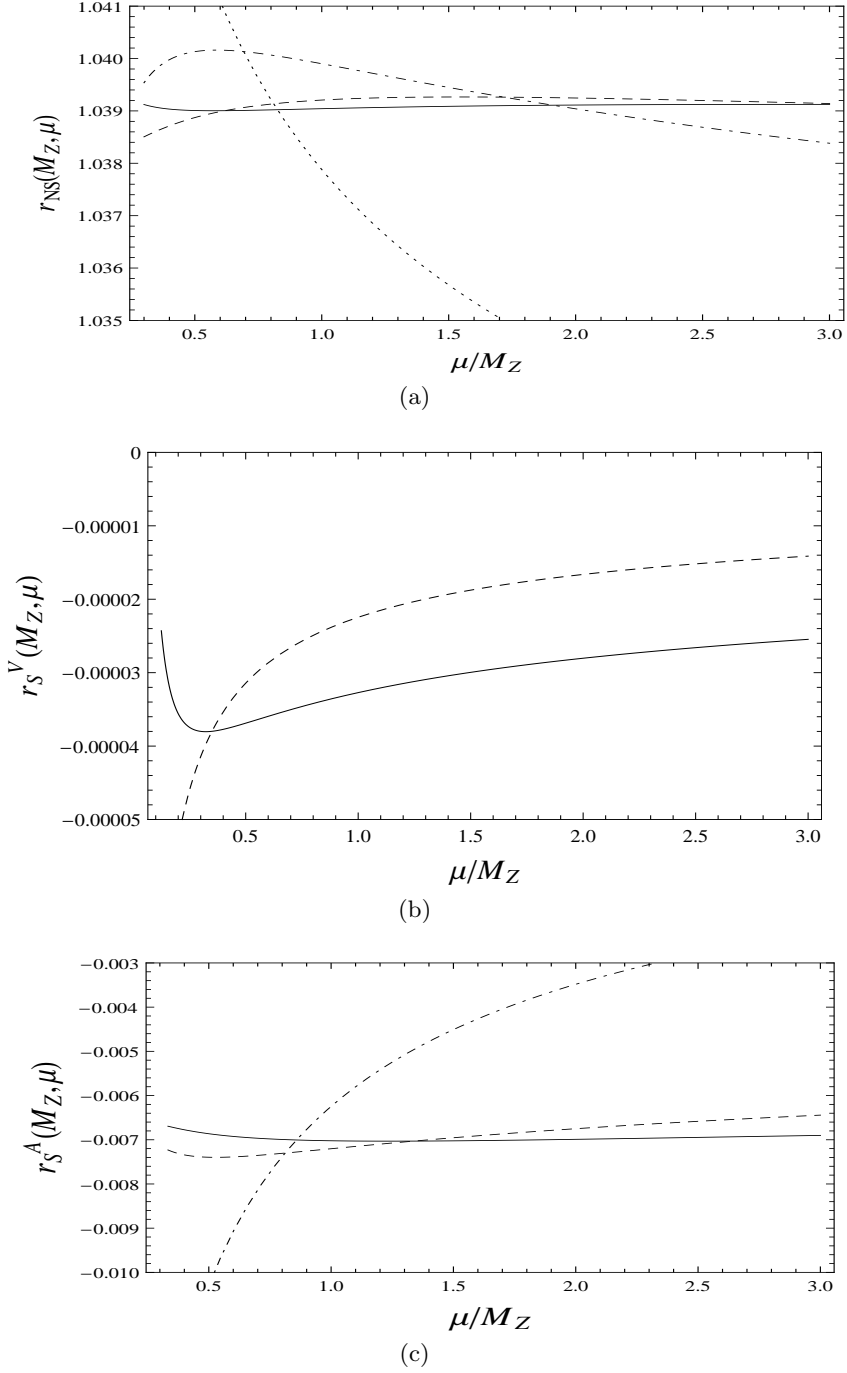

\centering
\subfloat[]{
\hspace{.45cm}
\includegraphics[width=.85\linewidth,height=5.0cm]{1201.5804.fig2a.ps}
\label{rns}
} \\
\subfloat[]{
\includegraphics[width=0.85\linewidth,height=5.9cm]{1201.5804.fig2b.ps}
\label{rsv}
} \\
\subfloat[]{
 \hspace{.2cm}
\includegraphics[width=.85\linewidth,height=5.0cm]{1201.5804.fig2c.ps}
\label{rsa}
}
\caption{
Scale dependence of (a) non-singlet $r_{NS}$ (b) vector singlet $r^V_S$ and (c) axial vector
singlet $r^A_{S;t,b}$. Dotted, dash-dotted, dashed and solid curves refer to ${\cal O}(\alpha_s)$
up to
${\cal O}(\alpha_s^4)$ predictions. $\alpha_s(M_Z)=0.1190$ and $n_l=5$ is adopted in all these
curves.
}
\label{Plots}
\end{figure}
}
%\clearpage

\noindent 
with $a_s=\alpha_s(M_Z)/\pi$ and $l_t=\ln (M_Z^2/M_t^2)$. Using for the pole mass $M_t$ the value 
172~GeV, the axial singlet contribution in numerical form is given by
\begin{equation}
%r^A_{S;t,b} = ...
%eq.4
% from 1201.5804
 r_{\rm S;t,b}^A = -4.3524\,\as^2 - 17.6245\,\as^3 + 87.5520\,\as^4~.
\end{equation}
The significant decrease of the scale dependence is evident from Fig.2.

%caption Fig2:
%Scale dependence of (a) non-singlet $r_{NS}$ (b) vector singlet $r^V_S$ and (c) axial vector 
%singlet $r^A_{S;t,b}$. Dotted, dash-dotted, dashed and solid curves refer to ${\cal O}(\alpha_s)$ 
%up to
%${\cal O}(\alpha_s^4)$ predictions. $\alpha_s(M_Z)=0.1190$ and $n_l=5$ is adopted in all these 
%curves.

Let us recall the basic aspects of these results:

\begin{itemize}
\item
The non-singlet term dominates all different channels. It starts in Born approximation and is 
identical for $\tau$ decay, for
$\sigma(e^+e^-\to hadrons)$ through the vector current (virtual photon) and for $\Gamma (Z\to 
hadrons)$ through vector and axial current.
\item
The singlet axial term starts in order $\alpha_s^2$, is present in
$Z\to hadrons$ and depends on $\ln M_Z^2/M_t^2$. Is origin is the strong imbalance between the 
masses of top and bottom quark \cite{Kniehl:1989bb}.
\item
The singlet vector term is present both in $\gamma^*\to hadrons$ and
$Z\to hadrons$ and starts in ${\cal O}(\alpha_s^3)$.
\item
All three terms are known up to order $\alpha_s^4$ and the total rate is remarkably stable under 
scale variations.
\end{itemize}

\section{Five-loop $\beta$ function}

Asymptotic freedom, manifest by a decreasing coupling with increasing energy, can be considered as 
the basic prediction of nonabelian gauge theories and was crucial for establishing QCD as the 
theory of strong interactions. The dominant, leading order prediction \cite{Gross:1973id,Politzer:1973fx} 
was quickly followed by the corresponding two- \cite{Caswell:1974gg,Jones:1974mm} and three-loop 
\cite{Tarasov:1980au,Larin:1993tp} results. Subsequently it took more than 15 
years until the four-loop result was evaluated \cite{vanRitbergen:1997va} and another seven 
years until this result was confirmed by an independent calculation \cite{Czakon:2004bu}. Now, finally, 
the five-loop result for QCD became available \cite{Baikov:2016tgj}, quickly confirmed and 
generalized to an arbitrary gauge group \cite{Herzog:2017ohr,Luthe:2017ttg,Chetyrkin:2017bjc}.

There are several reasons to push the QCD $\beta$-function to an order as high as possible. From 
the practical side it is important to compare experiment and theory prediction with the best 
achievable precision. From the theoretical side one expects that the perturbative series at some 
point starts to demonstrate its asymptotic divergence, shown by significantly increasing terms. 
However, as shown below, even up to fifth order the series exhibits a remarkably smooth behaviour 
with continuously decreasing perturbative coefficients. Let us, in a first step, recall the 
coefficients of the QCD $\beta$-function defined by
\begin{equation}
\beta(a_s)=\mu^2 \frac{d}{d\mu^2} a_s(\mu) =
-\sum_{i\ge 0} \beta_i a_s^{i+2}
{}.
\end{equation}
Using the same tools as those discussed in \cite{Baikov:2008jh,Baikov:2014qja} 
the $\beta$-function in fifth order is given by
%\begin{equation}
%full result of QCD beta function
%\end{equation}
%from 1606.08659
\begin{eqnarray}
 \beta_0 & = & \frac{1}{4}\Biggl\{ 11 - \frac{2}{3} n_f,  \Biggr\}, \ \ 
\nonumber
\beta_1    =  \frac{1}{4^2}\Biggl\{ 102 - \frac{38}{3} n_f\Biggr\}, 
\\
 \beta_2   &=&   \frac{1}{4^3}\Biggl\{\frac{2857}{2} - \frac{5033}{18} n_f + \frac{325}{54}
  n_f^2\Biggr\}, 
\nonumber
 \\
 \beta_3 & = &  \frac{1}{4^4}\Biggl\{ 
\frac{149753}{6} + 3564 { \zeta_3} 
        - \left[ \frac{1078361}{162} + \frac{6508}{27} { { \zeta_3}} \right] n_f
\nonumber
   \\ & & \hspace{15mm}
       + \left[ \frac{50065}{162} + \frac{6472}{81} { \zeta_3} \right] n_f^2
       +  \frac{1093}{729}  n_f^3\Biggr\}
{},\nonumber
%\ice{
\end{eqnarray}
\begin{eqnarray}
%}
%\\
&{}&\hspace{-4.5mm}\beta_{4} =  \frac{1}{4^5}\,\Biggl\{
\frac{8157455}{16} 
+\frac{621885}{2}  \sbz \zeta_{3}
-\frac{88209}{2}  \sbz \zeta_{4}
-288090  \sbz \zeta_{5}
%zero == 0
\nonumber\\
&{+}& \, n_f
\,\, 
\left[
-\frac{336460813}{1944} 
-\frac{4811164}{81}  \sbz \zeta_{3}
%\right.
%\nnb
%\\
%&{}& \hspace{40mm}
%\left.
+\frac{33935}{6}  \sbz \zeta_{4}
+\frac{1358995}{27}  \sbz \zeta_{5}
%zero == 0
\right]
\nonumber\\
&{+}& \, n_f^2
\,\,
\left[
\frac{25960913}{1944} 
+\frac{698531}{81}  \sbz \zeta_{3}
-\frac{10526}{9}  \sbz \zeta_{4}
-\frac{381760}{81}  \sbz \zeta_{5}
%zero == 0
\right]
\nonumber\\
&{+}& \, n_f^3
\,\,
\left[
-\frac{630559}{5832} 
-\frac{48722}{243}  \sbz \zeta_{3}
+\frac{1618}{27}  \sbz \zeta_{4}
+\frac{460}{9}  \sbz \zeta_{5}
%zero == 0
\right]
\nonumber
\\
&{+}&  \, n_f^4\,\,
\left[
\frac{1205}{2916} 
-\frac{152}{81}  \sbz \zeta_{3}
%zero == 0
\right]
\Biggr\}
\nonumber
%\label{beta_5l}
{}.
\end{eqnarray}

This result has, in the meantime, been confirmed in \cite{Herzog:2017ohr} and even 
extended to an arbitrary, simple, compact Lie group. The surprising pattern of the delayed 
appearance of higher transcendentals, already observed in lower orders, repeats itself in the 
present case: The transcendental numbers $\zeta_6$ and $\zeta_7$ that could be present in $\beta_4$ 
in principle, are evidently absent, similarly to the absence of $\zeta_4$ and $\zeta_5$ in the 
result for $\beta_3$.

Let us reemphasize the surprising smallness of the perturbative coefficients, characterized by the 
small deviations from the leading order result. Consider the ratio
$\bar \beta \equiv \frac{\beta}{-\beta_0 a_s^2} =
1+\sum_{i\ge 1}\bar\beta_i a_i'$
for two characteristic values of $n_f$:
%\begin{equation}
%\bar \beta(n_f=4)= ....
%\bar \beta(n_f=5)= ....
%\end{equation}
%from 1606.08659
\bea
\nnb 
\ovl{\beta}(n_f=4) &=&
1 +1.54 \,a_s+3.05  \,a_s^2+15.07  \,a_s^3+27.33 \,a_s^4,
\\ \nnb \ovl{\beta}(n_f=5) &=& 
1 +1.26 \,a_s+1.47 \,a_s^2+9.83 \,a_s^3+7.88 \,a_s^4
{}.
\eea
Indeed an extremely modest growth of the perturbative coefficients is observed. Remarkably enough, 
the rough pattern of the coefficients is indeed in qualitative agreement with the expectations for 
the $n_f$ dependence of $\beta_4$ based on the method of ''Asymptotic Pad\'e Approximant''  \cite{Ellis:1997sb}
(the boxed term was used as input):
\begin{eqnarray}
\beta^{APAP}_4  &\approx &  
%from 1606.08659
740 - 213\, n_f + 20\, n_f^2  -0.0486\, n_f^3   - \fbox{$ 0.001799 \,n_f^4$}
       ,
\nnb
\\
 \beta^{exact}_4 & \approx & 524.56 - 181.8 \, n_f + 17.16 \, n_f^2 
%%%          \nnb
%%% \\ &{}& \hspace{1cm}    
%%%
   - \,\,0.22586 \, n_f^3 -  0.001799 \, n_f^4
\nnb
{}.
\label{beta4N}
\end{eqnarray}
However, large cancellations occur for  $n_f=3$,4,5, leading to drastic disagreement for 
the final  predicitions for the corresponding values of  $\beta_4$.   

As stated before, the smallness of the higher order coefficients, in particular for the 
$n_f$-values of interest, leads to a remarkable stabilization of the results. The excellent 
agreement between $\alpha_s$ values from vastly different energy scales indeed persists in higher 
orders. Let us, as a typical example, recall the comparison between the strong coupling at the 
scales of $m_\tau$ and $M_Z$. Starting with the value $\alpha_s(m_\tau)=0.33\pm 0.014$ one arrives, 
after running and matching at the charm and bottom threshold at the value $\alpha_s^{(5)}=
0.1198\pm 0.0015$. From the direct measurement of $Z$-boson decays combined in the electroweak 
precision data, on the other hand, one obtains the result $\alpha_s^{(5)}=0.1197\pm 0.0028$ in 
remarkable agreement with the previous value.

\section{Summary}

A sizable number of four- and five-loop QCD results has been evaluated during the past years. 
${\cal O} (\alpha_s^4)$ corrections of Higgs boson decays to fermions, of $\tau$-lepton decays to 
hadrons, $Z$ decays to hadrons and  of corrections to the familiar $R$ ratio (with
$R\equiv \sigma(e^+e^- \to hadrons) / \sigma(e^+e^- \to \mu^+\mu^-)$) are among the most prominent 
examples. These calculations have been complemented by the most recent result along the same lines, 
the five-loop QCD $\beta$ function. No sign of an onset of the asymptotically expected
divergence of the series is observed. Excellent agreement between theory and experiment for a large 
number of predictions is observed. For the moment the precision of the theoretical prediction is 
significantly ahead of the experimental results.

The work of P.A.~Baikov is supported in part by the grant NSh-7989.2016.2 of
the President of Russian Federation and by the grant RFBR 17-02-00175A of the
Russian Foundation for Basic Research.  The work by K. G. Chetykin was
supported by the Deutsche Forschungsgemeinschaft through CH1479/1-1 and by the
German Federal Ministry for Education and Research BMBF through Grant
No. 05H15GUCC1.

\providecommand{\href}[2]{#2}\begingroup\raggedright\endgroup

\end{document}